\documentclass[twoside]{article}
\usepackage{graphicx}
\usepackage{array}
\usepackage{lipsum} 
\usepackage[sc]{mathpazo} 
\usepackage[T1]{fontenc} 
\usepackage{amsmath}
\usepackage{microtype} 
\usepackage[hmarginratio=1:1,top=32mm,columnsep=20pt]{geometry} 
\usepackage{multicol} 
\usepackage[hang, small,labelfont=bf,up,textfont=it,up]{caption} 
\usepackage{booktabs} 
\usepackage{float} 
\usepackage{hyperref} 
\usepackage{lettrine} 
\usepackage{paralist} 
\usepackage{abstract}

\usepackage{titlesec} 
\usepackage{fancyhdr} 
\title{\vspace{-15mm}\fontsize{24pt}{10pt}\selectfont\textbf{ Pedestrian Flow Simulation\\
Validation and Verification Techniques} }

\author{
\large
\textsc{Mohamed H. Dridi}\\
\normalsize Institute of Theoretical Physics 1 \\ 
University of Stuttgart \\
\normalsize \href{mailto:mohamed.dridi@itp1.uni-stuttgart.de}{mohamed.dridi@itp1.uni-stuttgart.de} 
\vspace{5mm}}

\date{27/09/2014} 

\begin{document}
\maketitle 
\thispagestyle{fancy}

\begin{abstract}
\noindent For the verification and validation of microscopic simulation models of pedestrian flow, we have performed experiments for different kind of facilities and sites where most conflicts and congestion happens e.g. corridors, narrow passages, and crosswalks. The validity of the model should compare the experimental conditions and simulation results with video recording carried out in the same condition like in real life e.g. pedestrian flux and density distributions. The strategy in this technique is to achieve a certain amount of accuracy required in the simulation model. This method is good at detecting the critical points in the pedestrians walking areas.   
For the calibration of suitable models we use the results obtained from analysing the video recordings in Hajj 2009 and these results can be used to check the design sections of pedestrian facilities and exits.
As practical examples, we present  the simulation of pilgrim streams on the Jamarat bridge (see fig. \ref{fig:Djamarat2}).

The objectives of this study are twofold: first, to show through verification and validation that simulation tools can be used to reproduce realistic scenarios, and second, gather data for accurate predictions for designers and decision makers.

\end{abstract}

\begin{multicols}{2} 

\section{Introduction}

\lettrine[nindent=0em,lines=3]{I} n this paper we attempt to explore the methods that can be used to make the results made by a software or simulation tool more authentic or believable. A set of statistical data taken from real life experience can be used to check the output values created by the simulation tools to validate the simulation model. This method is referred to as statistical technique method and can be applied to simulation models, depending on which real-life data is available \cite{Dijkum1998}. In general lack of empirical data makes the verification of any simulation model a complicated task.

In case the real data are not available - the simulation data obtained by the simulation tools 
are still guided by the condition of a statistical theory and probability distributions on the design of experiments \cite{Knepell1993}.

In case only output data is available - the values carried by the simulation model can be compared with well-known statistical data \cite{Knepell1993}. 

If data can be collected on both system input and output trace-driven simulation becomes possible, model validation can be done through comparing the collected data with the simulation results. In trace-driven simulation, the simulation input data are identified by the trace data collected by a myriad of instruments and methods \cite{Knepell1993}.

What, however, does 'validation' mean? The term validation will be used to refer to various processes.
The process of examining whether the acceptability and credibility of the conceptual model is referred to as validation, it is an accurate method to check the actual system being analysed. Validation can help to develop the right model. Verification is a process to check simulation output for acceptability and controlling whether the results made by the computer program are compatible with the real data collected about the same system \cite{Knepell1993}.

Concerning this topic many books could be written to describe the philosophical and practical aspects involved in validation (see, the monograph by Knepell, and Arangno 1993)\cite{Knepell1993}. For this reason, I identify validation as systematic examination of the simulation model whether (if) it displays or illustrates the real world in a reasonable time, either as a procedure to check for correctness or meaningfulness of the resulting data. Validation is a process to check the ability of the model to reproduce the real system. In the next sections we will concentrate on validation that uses mathematical statistics and comparison with video recordings of real situations.

\begin{figure}[H]
\begin{center}
\includegraphics[width=\columnwidth]{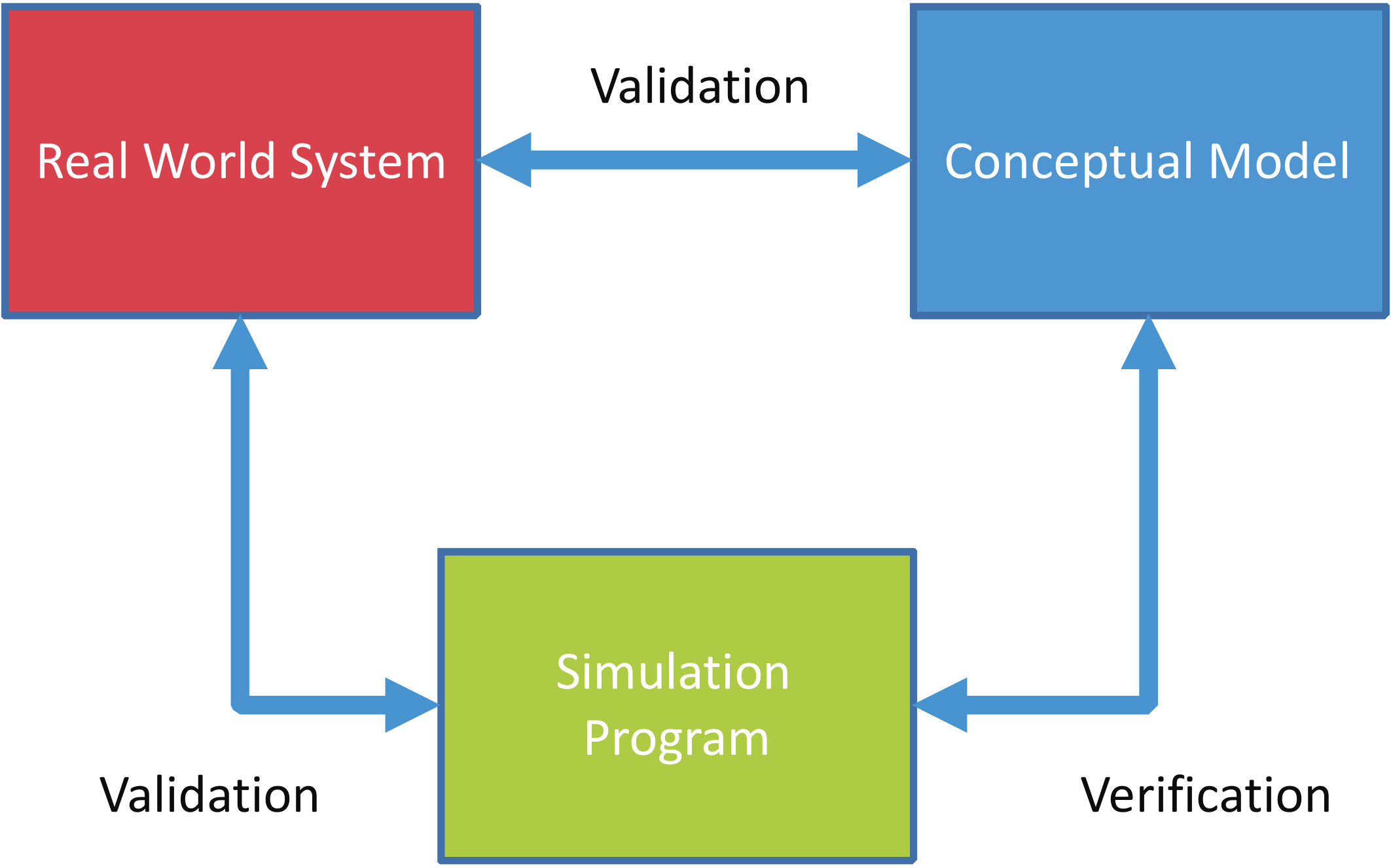}
\end{center}
\caption{Validation procedure for a given simulation model.}
\label{fig:validation-2}
\end{figure}

Since modelling and simulating has become very important in many domains in modern science, much literature on verification and validation of a simulation models have appeared: see the web (http://manta.cs.vt.edu/biblio/), and the detailed surveys in
Beck et al. (1997)\cite{Beck1997}, Kleijnen (1995b)\cite{Kleijnen1995b}, and Sargent (1996)\cite{Sargent1996}. Important work concerning the choice of statistical tests to validate a model was made by Kleijnen (1999)\cite{Kleijnen1999}.

For the first step we try to compare our video taking with the simulation results. A lot of phenomena (like the lane formation, oscillation effect and edge effect) can be seen, to make sure if our simulation reproduces a part of the reality. For this investigation we need to make scenarios for the next video observation in the great mosque in Mecca.

\section{Validation}
\subsection{Verification, validation and testing techniques}
This paragraph describes different validation techniques and tests, used in model examination and validation. Most techniques described here are found in literature, although some may be described slightly differently. They can be used either subjectively or objectively. In the "objectively" case we attempt to implement mathematical methods using a kind of statistical test e.g.\ confidence intervals and hypothesis testing. A combination of techniques is generally used. These techniques are used for the examination and validation of sub-models and the universal model \cite{Sargent2005}.

\begin{itemize}
\item 	Comparison to other models: 
In a verification and validation of a simulation model process, different results (e.g. outputs) of the simulation model will be compared with the results of other models. For example, (1) comparison of a simple case of a simulation model with well-known results of empirical models, and (2) the comparison of the simulation model with other validated models with the same properties.
\item 	Event validity: 
The appearance of events in a simulation model will be compared with those of the real system to determine if they are identical.
\item 	Extreme condition tests: 
The model structure and outputs should be credible for any extreme and improbable combination of levels of factors in the system.
\item 	Face validity: 
Experts or specialists in the system will be asked about the suitability of the model and its behaviour. For example, is the logic in the conceptual model true and are the model input-output relationships appropriate.
\item 	Historical data validation: The system can take advantage of the historical collected data to calibrate itself, specifically the data collected on a system for building and examining the model, a part of the data can be used to establish the model and the remaining data is used to determine whether the model behaves as the system does. (This testing is led by driving the simulation model samples from the distributions or traces) \cite{Balci1982a, Balci1982b, Balci1984b}.
\item 	Multi stage validation: 
Another efficient method for validation a simulation tool was proposed by Naylor and Finger\ (1967)\cite{Naylor1967}. It consists in associating three well known methods of rationalism, empiricism, and positive economics into a multi-stage process of validation. This technique is based on (1) evaluation and development of the simulation model with respect to the theory, observations, and practical experience, (2) validating the model using possible existing empirical data, and (3) comparing the results (output) made by the simulation model with the real system. 
\item 	Operational graphics: Measured values of various performances e.g.\ using statistics for time series, are illustrated graphically while the model runs over time; i.e. a visual indicator of performance shows how the program behaves during run time to ensure the correct performance of the simulation tools.
\item 	Sensitivity analysis: A sensitivity analysis is a powerful technique for validating systems.
This validation method consists in changing the input parameters of the simulation or internal parameters of the model to realize how the model's output will be affected. If the system does the right things, the same relationships resulting from the model should be visible in the real system.
Using this technique both qualitative (directions only of outputs) and quantitative (both directions and exact amount of outputs) properties of the system can be verified. 
Parameters cause important changes in the behaviour of the model (sensitive parameters). These parameters have a high importance for the model and the simulation results (this may require iterations in model development).
\end{itemize}

\section{Calibration and validation of PedFlow model}
\label{PedFlow-Calibratioin}
In this section we present a variation of different techniques used to calibrate and validate the PedFlow simulation model.
PedFlow is a microscopic simulation model, which was developed by L\"{o}hner Simulation Technologies International, Inc. (LSTI) \cite{Loehner}.
For verification and validation, data was provided by the Institute of
Hajj research and the Ministry of Hajj, consisting of layout information, pilgrim numbers, and Hajj schedules. We augmented this data with camera-based
observations at several stairways, gates and the piazza inside and outside the Great Mosque in Mecca. This collected data can be used as input parameters of the simulation and improves the acceptability and accuracy of the data carried by the simulation. 

PedFlow must model all processes that are related to pedestrians inside and outside the Haram at the normal and the busiest rush hours of the Hajj events such as: walking, performing activities, and route choice.
In order to validate pedestrian flow modelling in PedFlow and to study pedestrian traffic flow movement during the Hajj in detail, observations were collected on the Haram in Mecca during the Hajj 2009. These observations concerned the Tawaf, Sa'y, (individual) walking times, and other sites such as the Haram gates before and after each prayer.
These observations are very helpful in obtaining the data that will be used to verify our simulation tool PedFlow. This data concerns the numbers of pilgrims going in and out of the Haram and individual walking times and densities of pedestrians on the Mataf. Finally, a comparison is made between the observations and modelling results of PedFlow, in order to check the validity of PedFlow with respect to pedestrian traffic flow operations.

Since this investigation is concerned with studying safety and fluidity of large scale pilgrim flows at pilgrimage places in Mecca, the validation of the simulation tool is mainly concentrated  on pedestrian traffic flow at the holy places. The main variables to be observed and compared with the model predictions are:
\begin{itemize}
\item Walking speeds.
\begin{itemize}
\item On the stairs (upward and downward directions).
\item On the piazza and the Mataf of the Haram.
\end{itemize}
\item Densities over time and space.
\begin{itemize}
\item Video recording.
\item Fundamental diagrams Predtechenski and Milinski \cite{Predtetschenski-Milinski1969}.
\end{itemize}
\item Layout information
\begin{itemize}
\item Data about the boundary condition and environmental information.
\end{itemize}
\end{itemize}

\subsection{Validation through comparison with other models}
The credibility of the data produced by the pedestrian microscopic simulation model can be validated through comparing with results obtained by other models having the same characteristics, although we mention that the comparison with other simulation tools is necessary for the acceptance of the data but not sufficient.  
Different results (e.g. outputs) of the PedFlow simulation
model, being validated, are compared with results of other
models. For example, emergent lane
formation generated by many simulation models, e.g. Blue \cite{Blue2001}, who used a cellular automata model. Lane formation in bidirectional flow and clogging effects at bottlenecks in case of emergency situation were realized by Helbing, Molnar, and Vicsek \cite{Helbing-Farkas2000a}, who use a social force model.
First a simple case of a simulation model is compared with known results of analytic models \cite{Predtetschenski-Milinski1969}, and second the simulation model is
compared with other simulation models that have been validated, such as social-force models (see \cite{Helbing-Farkas2002} and the references therein) and cellular automata, e.g \cite{Fukui1999, Muramatsu2000}.

\subsubsection{Walking through a narrow passage;}
\begin{figure*}[ht]
\begin{center}
\includegraphics[width=1.0\linewidth]{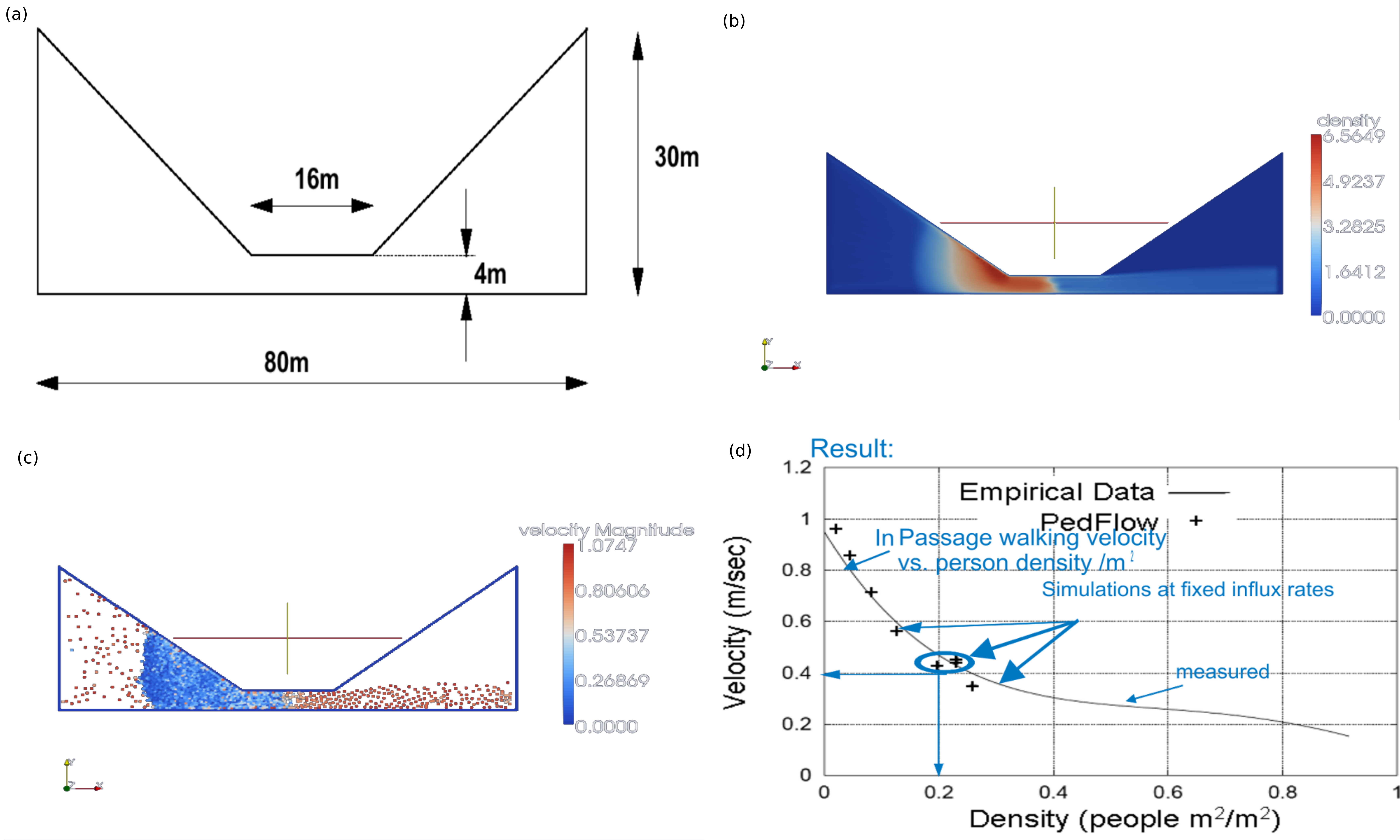}
\end{center}
\caption{This figures illustrates a pedestrian flow walking through a narrow passage: (a) bottleneck geometry; (b) the density index; the density map illustrates how the pedestrian density rises in the congested area. Red color indicates high density which can reach 7 people/m$^{2}$, while blue color indicates low pedestrians density; (c) the velocity index, the blue color indicates the lowest velocity; (d) decrease of the pedestrian velocity in the passage area as a function of the local density, the measured data are represented by the crosses in the graph, they are consistent with the empirical data of Predtetschenski and Milinski  \cite{Predtetschenski-Milinski1969} represented by the solid line.}
\label{fig:bottleneck1}
\end{figure*}

Our first set of simulations consisted of pedestrian flow through a hallway with a narrow passage (see fig. \ref{fig:bottleneck1}) (a). The hallway was 80 m wide and the
narrow passage was 16 m long and 4 m at the narrowest point. Each
pedestrian's desired speed was set at a walking speed for adult pedestrians in normal conditions $v_{d}$ = 1 $\pm$ 0.02 m/s; 
relaxation time: $\tau$ = 0.50 $\pm$ 0.1 s; (smaller times $\rightarrow$ more aggressive) and
pedestrian radius: $R$ = 0.2 $\pm$ 0.02 m; (smaller radius $\rightarrow$ smaller repulsive forces).
Repulsive potentials were assumed to decrease exponentially.
The relaxation time is the time needed to reach 90 percent of the desired velocity.

\begin{figure}[H]
\begin{center}
\includegraphics[width=\columnwidth]{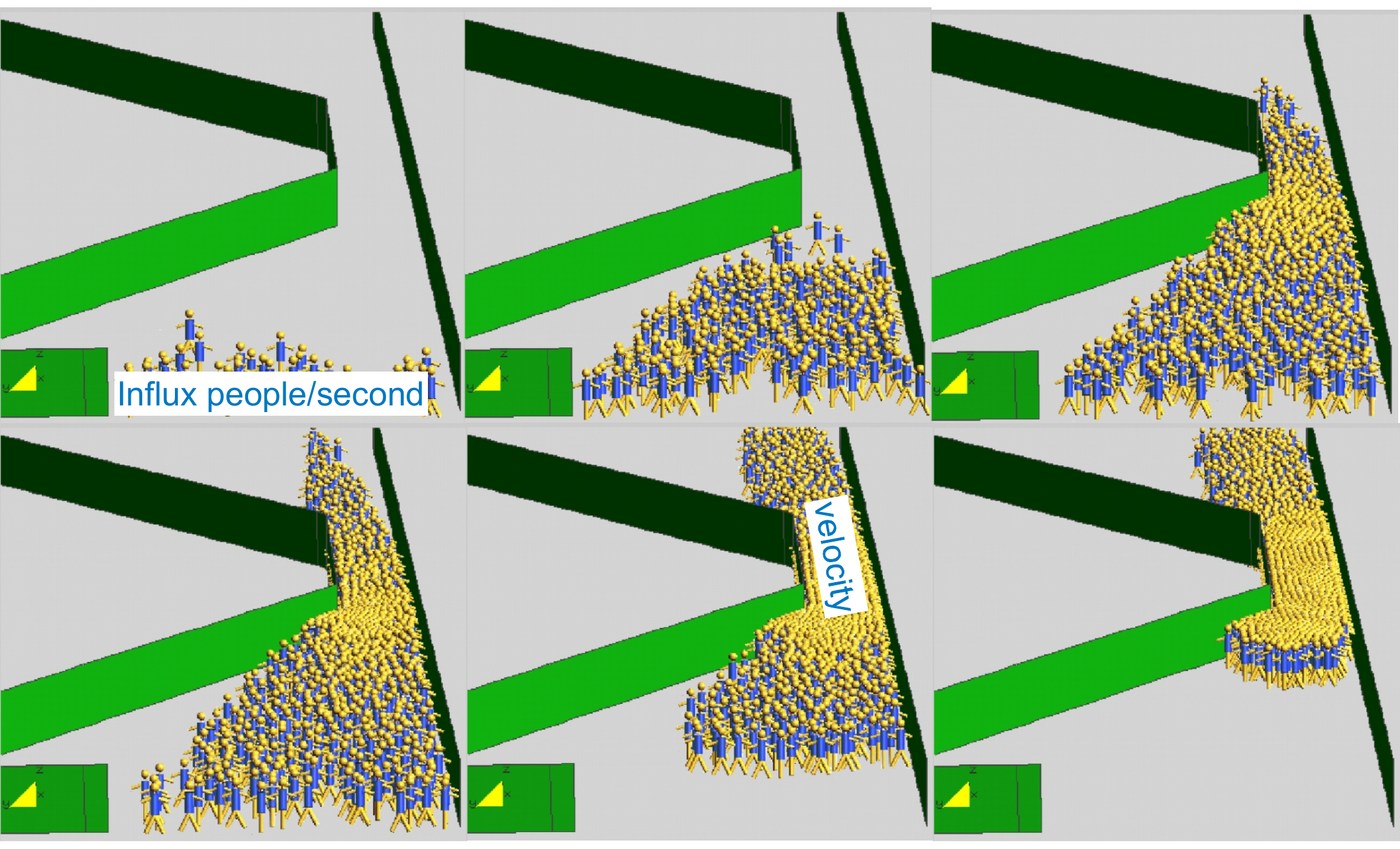}
\end{center}
\caption{Step by step simulation of a pedestrian crowd walking through a passage (PedFlow animation results).}
\label{fig:walkingpassage-1}
\end{figure}

This experiment is realized with constant influx, that means if a pedestrian has passed through the passage, he will be replaced by a new pedestrian at a random starting location, i.e. at the
entrance of the hallway to keep the  number of people in the hallway constant. The mean velocities (measured in the passage area) for different pedestrian influxes and the results are illustrated in figure \ref{fig:bottleneck1} (d). These results are consistent with those obtained by Predtetschenski and Milinski \cite{Predtetschenski-Milinski1969} and other fundamental diagrams, who also found out reduced velocities due to the tendency of pedestrians to converge at the same time in the direction of the passage area when the
hallway is narrow. This caused blocking and the velocities to drop and the density to increase with time, creating bottlenecks and clogging, (see fig. \ref{fig:walkingpassage-1}).

Pedestrian motion in passages is one of the few cases where reliable empirical data exists. In order to assess the validity of the proposed pedestrian motion model, a typical passage flow was selected. The geometry of the problem is shown in figure \ref{fig:bottleneck1} (a). Pedestrians enter the domain from the left and exit to the right. In this case, each pedestrian has the goal of first reaching the entrance of the passage, then traversing it to the other end, and finally to exit on the right. Typical snapshots during one of the simulations are shown in figure \ref{fig:walkingpassage-1}. 

The resulting data of the simulation are shown as crosses in figure\ref{fig:bottleneck1} (d). The data flows are shown in a graph besides the analytical data from Predtetschenski and Milinski. This graph corresponds to specific parameters and illustrates a defined simulation state, although they exhibit the relation between the input parameters and the simulation results. In the low density range the data are synchronized with a high accuracy. There are no deviations of the simulation values and the analytically data. The walking speed drops in dependency of density.     The small deviation in the start-velocity, can be traced back to the input parameters. 

\subsubsection{Example: simulation of the Tawaf movement}
In this test we try to verify the simulation response by running a simplified version of the simulation program with a known analytical result. If the output data resulting from the simulation model do not exhibit a significant deviation from the known empirical data, this result can then be used to validate the model.

\begin{figure*}[ht]
\begin{center}
\includegraphics[width=1.0\linewidth]{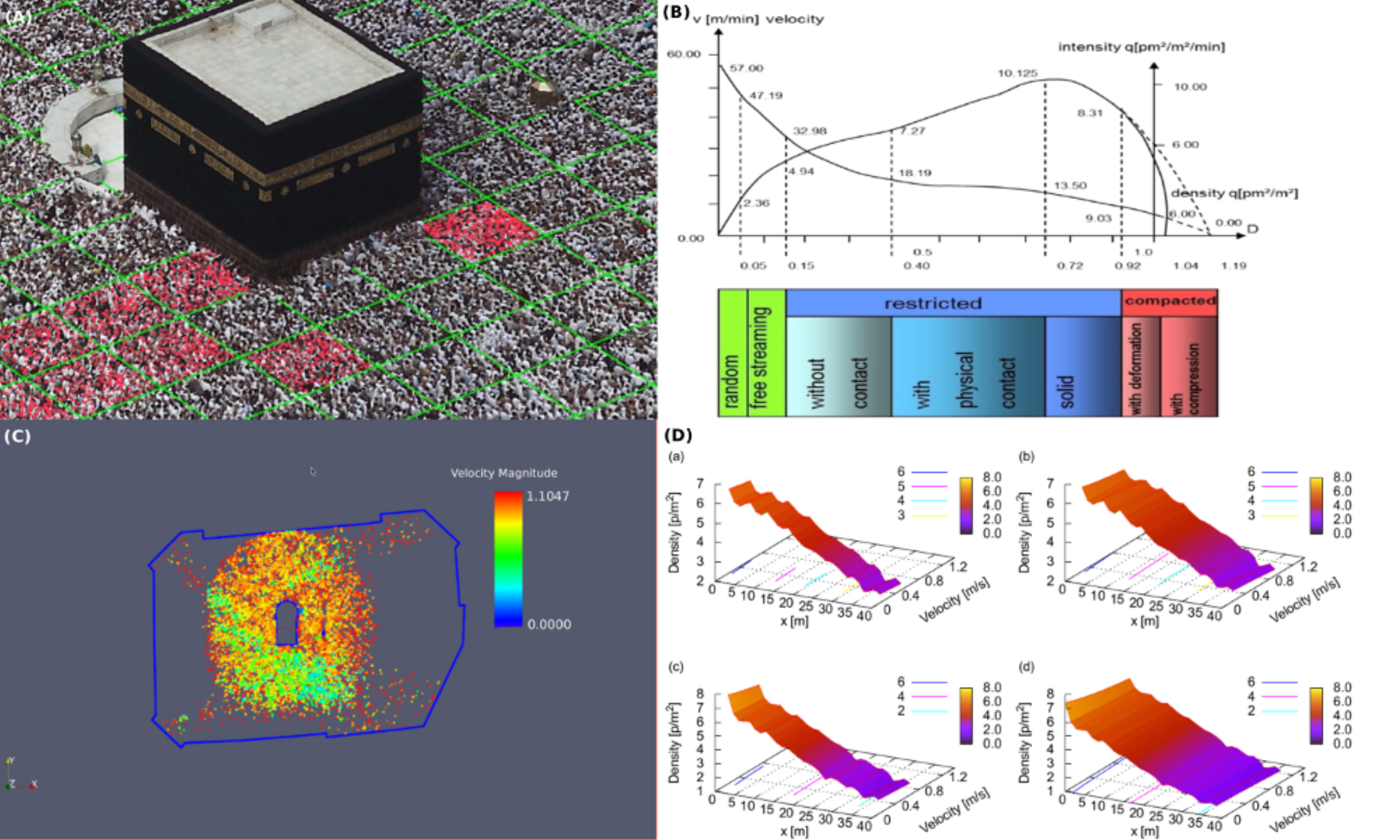}
\end{center}
\caption{Tawaf movement simulation. (A) This figure illustrates the Mataf area at rush hour divided in regular cells. The number of pedestrians in every cell as a function of time is determined through repeating the counting process many times. The average value is identified as local density $\rho(\vec{r}, t)$; (B) velocity-density diagram: Empirical relation between density and velocity according to Milinski and Predtetschenski \cite{Predtetschenski-Milinski1969}. The partition refers to domains with qualitatively different decrease of the velocity; (C) pilgrims movement simulation within the Mataf area, the red color indicates the desired velocity 0.9 to 1 m/s while turquoise color the lowest; (D) velocity-density distribution as a function of the distance from the Kabaa wall, the curves (b, d) illustrate the simulation results while the curves (a, c) show the results obtained by a calculation according to the Predtetschenski and Milinski fundamental diagram.}
\label{verification2}
\end{figure*}

Through the simulation of pedestrian flow on the well known geometry of the Mataf in the Haram Mosque (see fig. \ref{verification2}) we intended to check simulation output for credibility. 
We performed different simulation runs  for several input scenarios and tested whether the output is reasonable. It is easy to compare certain performed measurements with other computed results.
Using animation is another method to improve the simulation model. The resulting data of the simulated system is displayed in a series of snapshots of the animation of the model users.
Since the model developers and model users are familiar with the real system, they can ameliorate the performance of the program and detect programming and conceptual errors.

Figure \ref{verification2} (C) illustrates typical simulation results for Tawaf movement (circling the Kaaba seven times in  a counter-clockwise direction). The entire influx consists of three one-directional pedestrian flows coming from three entrances. The velocity indicator shows that the movement in the edges of the Mataf area is faster than in the area of the Kaaba. The picture shows a snapshot of the simulation, which has a particularly high maximum density
of 6.5 persons/m$^{2}$. Note that the pedestrian density is very high at the places where the Tawaf begins and ends, and the clumping of pedestrians going in opposite directions, when the pilgrims finish the Tawaf. The average density for many runs was 5 to 7 persons/m$^{2}$. Figures \ref{verification2} (D; a) and (D; c) illustrate the velocity-density distribution on the Mataf area during the rush hour calculated according to Predtetschenski and Milinski. This result agrees well with the measured data shown in Figures 4 (D; b) and (D; d).

\subsubsection{Example: Al-Jamarat Bridge}
The simulation of high density pedestrian flow streaming the Jamarat area during the rush hour of the Hajj period revealed a great technical progress in the modelling, simulation and better understanding of how large crowds alter. 
In the past many fatal accidents happened in this extremely dangerous area, where a large number of pilgrims stream through the site and try to stone 
the pillars in a relatively short period of time. Since the movement of pilgrims is very slow an accumulation effect on both sites of the pillars arises. This leads to physical jamming, pilgrims trampling, and in the worst situation to the death of pedestrians underfoot.
To accomplish the safety of millions of pilgrims walking this overcrowded area every year and for better fluidity of pedestrian flow near the pillars, the proposal was made to build a bridge with a 5-level structure to ease the process of performing this ritual. The bridge was designed to satisfy the international standard criterion of pedestrians safety, especially during overcrowding, and this concept arose from the idea to conduct the pilgrims flow in one direction without any counter flow.
The Saudi government designated Professor Dr. Saad A. AlGadhi (expert in transportation management and design) and Dr. G. Keith Still (the crowd dynamics expert) to evaluate a model using crowd dynamic software tools to improve the conceptual design \cite{AlGadhi2003}. This study produced a lot of data and information about:

\begin{itemize} 
\item Sufficient arrival capacity 
\item Sufficient throwing area 
\item Sufficient space (density $\leq$ 4 Hajjis per square meter) 
\item  Sufficient passing area 
\item  Sufficient egress capacity
\end{itemize}

in the Jamarat bridge area that can be used to validate other pedestrian simulation tools. For example, published data about the Jamarat bridge capacity, in-flux and out-flux, demonstrate that the total available ingress width must be greater than 28 meters to allow 125,000 pilgrims per hour. This is a minimum requirement and provision for security forces/civil defence, bi-directional/counter flow 
and hesitation (pilgrims stopping to rest) where the longer ingress ramps have additional width requirements \cite{AlGadhi2003}.

\begin{figure*}[ht]
\begin{center}
\includegraphics[width=1.0\linewidth]{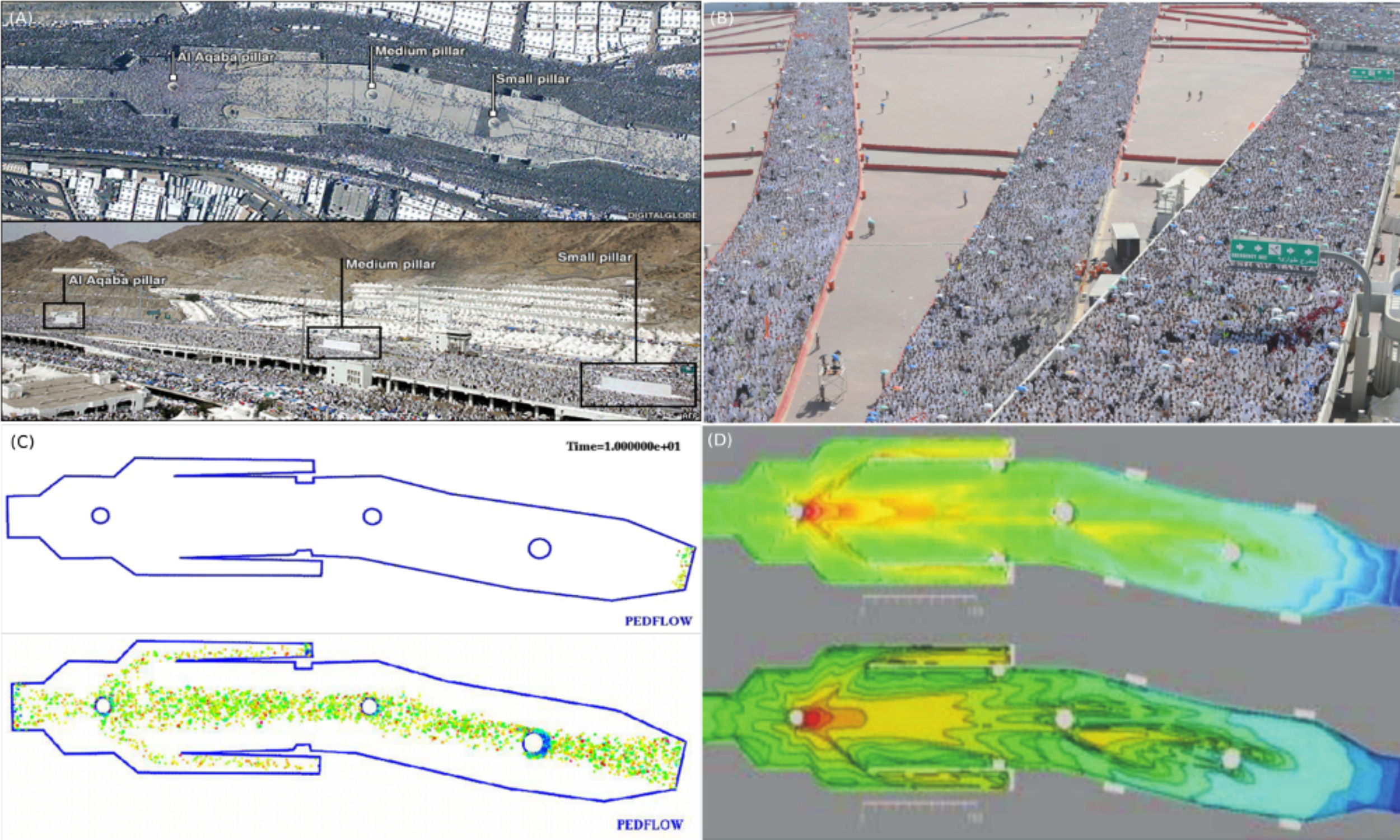}
\end{center}
\caption{During Hajj, pilgrims flock to the Jamarat Bridge in Mina to cast stones at three pillars representing the devil. The cylindrical pillars (A)(top) were replaced by short walls (A)(bottom) after a previous fatal stampede in 2004. The idea was to improve crowd flow and reduce congestion. 
(A) shows the geometry and the location of the stoning pillars;
(B) shows a huge number of pilgrims streaming toward the pillars;
(C) Al-Jamarat Bridge microscopic simulations/PedFlow and
(D) Microscopic simulations/Myriad \cite{Still2003} (red color means high density; yellow color means middle density; green and blue color means low density).}
\label{fig:Djamarat2}
\end{figure*}

Figure \ref{fig:Djamarat2} illustrates how a combination of microscopic and macroscopic techniques can assess the progression of queues approaching the Jamarat (above - Simulex/Myriad, below Myriad and site photograph)\cite{Still2003}.

The other set of data about the Jamarat bridge was published by Helbing after the onset of the crush event of 2006 in his paper: The Dynamics of Crowd Disasters: An Empirical Study 2007 \cite{Helbing2007}.
His video analysis revealed a lot of data and information about the average local speed, average local flows and the average local densities in the Jamarat area before and after the deathly crush accident. It was found that the pedestrian density near the pillars area can reach a huge value of 9 persons/m$^2$.

To assess the validity of the PedFlow simulation model and for improvements of resulting data we apply analytical and comparative tests. These tests are used to compare the simulation output with the output from other simulation tools e.g. Simulex/Myriad \cite{Still2003}. 
Compared to other models PedFlow is more sophisticated to predict high density crowd dynamics. The simulation result is shown in figure \ref{fig:Djamarat2} (C) and (D). Of course the simulation input takes advantage of the published data to predict accurate results.

\subsection{Validation through visualization and comparison with the real world}
The aim of most procedures and methods testing model validity is to determine the similarity between the results carried out by the conceptual model and the collected data. The better the simulation output resembles the output from the real system the better the results in general.
The animation and visualization of the output simulation data are necessary to prove the credibility of the system, moreover this test is very important to examine how close the data is to the real world.

\subsubsection{Crowd visualization}
A literature survey reveals several investigations and animation methods which have been proposed to provide more realism in the conceptual model simulating large scale pedestrian motion. Treuille and Shao \cite{Treuille2006, SHAO2007} suggested a method that increases the degree of accuracy and realism of crowd simulations. 
For example a realistic human like character is an essential role in the animation of high density crowd simulation. They illustrate the effects and interactions between the individuals itself within the crowd and the individuals and their environment. This yields a better prospect about the density distribution of pedestrians in a given site.

In the context of pedestrian animation we considered the technique of motion graphs \cite{Kovar2002} in order to provide advanced behavioural human characters. We attempted to modify the motion graph approach to associate an existing database of short MoCap (Motion Capture) animations into a larger clip of continuous motion.
However, in our approach the pedestrian movements are expressed as paths or trajectories of the character extracted from unlabelled motion capture data. This technique modifies the character's position and orientation for the entire animation clip. The trajectory and orientation of a character in a BVH (Biovision Hierarchical data) MoCap animation is interconnected, and one cannot be modified without influencing the other, hence rendering the animation unrealistic. Our technique allowed us to produce a continuous and longer sequence of animations using an existing database of MoCap animations and joining the animations together. Behaviour is closely related to the corresponding animation. 
It is this binding between behaviour and animation that we intended to utilize to validate our model.

\begin{figure*}[ht]
\begin{center}
\includegraphics[width=1.0\linewidth]{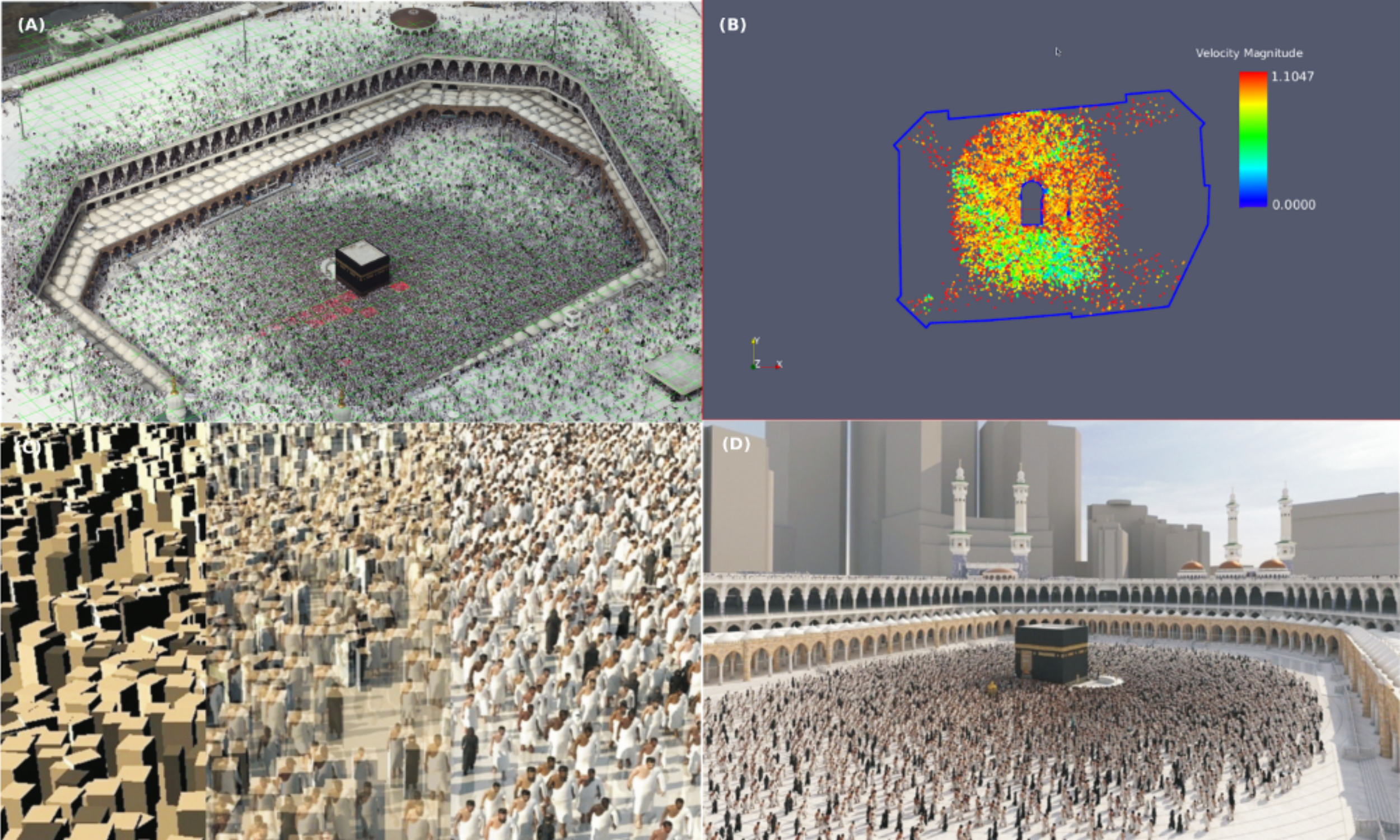}
\end{center}
\caption{Crowd visualisation, real vs. virtual world: (A) The real world represented in the Mataf top view; (B) simulation snapshot of the Tawaf movement; (C) crowd visualisation (characters implementation); (D) illustrates the reproduced virtual world.}
\label{fig:polygons-1}
\end{figure*}

The processes used to describe animated movement of one or more objects or persons are presented in this paragraph.
From tabular values carried by the microscopic simulation data results the path and velocity vectors of the agents are determined. The coordinates and velocity of every pedestrian at any time is given by the simulation data. The animation of the characters is designed in two steps: first we attach a polygon to every coordinate, and next we attach every polygon to a human character, (see fig. \ref{fig:polygons-1}(C)).
The motion tracking and motion animation is applied in many disciplines and scientific fields like entertainment, and medical applications, and for validation of computer vision \cite{Noonan2009}. However, we distinguish two types of animations: the  film and game-makers, who take advantage from this technique to reproduce multitudinous number of avatars. The second type of animation is based on exact simulation results and illustrates more realism, which can be used to help the decision maker to manage huge crowds, detect critical points in a closed area and to help the architects and designer to establish the number of fire exits required for a building. This animation can contribute to the validation of the simulation tools.

\subsubsection{Validation through comparison with the real world}
For validation of a simulation model, it is necessary to compare the simulation output with real-world data, such as video recordings representing the same circumstances of the simulation. This method can ascertain a lot of effects and behaviours that appear in crowds. 

Through observation of pilgrim flow we attempt to validate and verify the crowd dynamic model tool PedFlow.
The obtained real data presented in paper \cite{dridi2014} is used to verify a microscopic crowd dynamics model developed to solve  complicated problems concerning high density crowd behaviours. The crowd dynamics model attempts to simulate the global movement of each individual influenced by the temporal circumstances and the surrounding crowd. A good agreement between the predictions and observations will validate the prediction model. 

\section{Validation of PedFlow using optical flow method} 
\subsection{The optical flow method}  
In the last years optical flow is considered as one of the most important techniques concerning image processing and computer vision. Computing of optical flow vectors using consecutive image sequences is achieved in two different ways: gradient methods and correlation methods.
Many studies show that optical flow techniques can be successfully used to identify or recognize moving objects, e.g. moving cars or walking person,  \cite{Ricquebourg2000}. Compared with other models this approach is able to operate with relatively low computational expenditure or visibility requirements on a diversity of entities, permitting reconstruction of object trajectories with high accuracy  from video recordings. The detection of movement can be determined by the introduction of different sets of image sequences - by considering the different details between two images - since this is more accurate in computational calculation and prediction \cite{Masoud2001}. The difference in image brightness can then be analysed further to extract movement vectors that describe the motion of the drops (entity) captured in the respective images. This method is based on video segmentation and position identification, rather than motion recognition by analysing frame by frame sequences of ordered images. 

Over the last decades, computer scientists have worked in different ways to reconstruct the trajectory of moving objects. Many studies and investigations appearing in different scientific fields attempt to compute the optical flow given by a sequence of images (see the comprehensive surveys 
\cite{Barron1994, Beauchemin1995}). 
The gradient and correlation methods are the mainly used techniques for computing and calculation of optical flows. In addition to these, there are other statistical methods which are able to estimate the motion parameters \cite{Fan1996} and the use of phase information \cite{Fleet1990}. The approach proposed by Hayton establishes a relationship between optical flow and image registration techniques \cite{Hayton1999}. 

Let us denote by $I(x, y, t)$ the image intensity function associated with to the pixel grey value at location $(x, y)$ of the image at time $t$. Gradient-based techniques are  predicated on the intensity conservation assumption 
\begin{equation}
I(x, y, t) = I(x+ \delta x, y+ \delta y, t+ \delta t), 
\label{eq:opticalflow-1}
\end{equation}
which can be expanded in a Taylor series neglecting higher order terms \cite{Horn1981}. In general, 
gradient-based techniques are accurate only when the intensity is preserved,
and the Taylor series approximation stays reasonable when frame-to-frame displacements due to subjects motion are a part of a pixel. To reduce the errors resulting from using this technique and to compute flow vectors over a larger image region an iteration method is deployed. 

Correlation-based techniques will be useful if the image sequences do not meet the conditions required for gradient-based techniques, that means the brightness intensity is not preserved, for example in cloud \cite{Wu1995} and combustion \cite{Sun1996} images. Such techniques try to establish correspondences between invariant characteristics between frames. Typical features might be blobs, corners and edges \cite{Clocksin2000}.

\subsection{Motion analysis and object tracking}
As already mentioned optical flow is a method to estimate object motions through brightness intensity changes in sequences of consecutively ordered images. A brightness intensity region variation related to the average pixel intensity of each image in a sequence of crowd images is used to estimate the pedestrian density distribution at various sites.

The technique permitting pedestrian's movement capture e.g. extracting information about pedestrian speed, using video footage obtained from CCTV observation of urban crowd movement surveillance and image processing can be traced back to Velastin \cite{Velastin1994} and \cite{Velastin1993}, who use algorithms operating on pixel intensities under a certain condition (such as a high frame rate) \cite{Johnston2004}. Other techniques and methods to compute the optical flow regarding changes in pixel intensities in a series of images sequences are developed by \cite{Seki2000, Vannoorenberghe1996, Masoud2001, Yonemoto2003}. 

Optical flow is defined as an apparent motion of image brightness. Let $I(x,y,t)$ be the image brightness that changes in time to provide an image sequence. Two main assumptions are made:
\begin{itemize}
\item Brightness $I(x,y,t)$ smoothly depends on coordinates $x$, $y$ in a greater part of the image.
\item Brightness of every point of a moving or static object does not change in time.
\end{itemize}
Let some object in the image, or some point of an object, move and denote the object displacement after time $dt$ by $(dx, dy)$. Using Taylor series expansion for brightness $I(x,y,t)$:
\begin{multline}
I(x + dx, y + dy, t + dt) = I(x, y, t)\\ + \frac{\partial I}{\partial x} dx +
\frac{\partial I}{\partial y} dy + \frac{\partial I}{\partial t} dt + ......,
\end{multline}
where $.....$ are higher order terms,
then, according to assumption 2:
\begin{equation}
I(x + dx, y + dy, t + dt) = I(x, y, t), 
\label{eq:I}
\end{equation}
and
\begin{equation}
\frac{\partial I}{\partial x} dx + \frac{\partial I}{\partial y} dy + \frac{\partial I}{\partial t} dt + ......= 0,
\end{equation}
Dividing (\ref{eq:I}) by $dt$ and defining
\begin{equation}
\frac{dx}{dt} = u, \frac{dy}{dt} = v 
\end{equation}
results in 
\begin{equation}
-\frac{\partial I}{\partial t} = \frac{\partial I}{\partial x} u + \frac{\partial I}{\partial y} v,
\label{eq:2}
\end{equation}
usually called the optical flow constraint equation, where $(u, v)$ are components of the optical flow field vector in $x$ and $y$ coordinates respectively.

The movement recognition in this work is based on the optical flow method extracting data from picture sequences using the
Lucas and Kanade algorithm \cite{Lucas1981}.
It considers a group of adjacent pixels and supposes that all of them (the group of adjacent pixels) have the same velocity. It finds an approximate solution of the above equation (\ref{eq:2}) using the least-square method by solving a system of linear equations. The equations are usually weighted. 
Here the following $2\times2$ linear system is used:
\begin{multline}
\sum_{x,y} W(x, y)I_{x}I_{y}u  + \sum_{x,y} W(x, y)I_{y}^{2}v\\ = - \sum_{x,y} W(x, y)I_{y}I_{t},
\end{multline}
\begin{multline}
\sum_{x,y} W(x, y)I_{x}^{2}u  + \sum_{x,y} W(x, y)I_{y}I_{y}v\\ = - \sum_{x,y} W(x, y)I_{x}I_{t},
\end{multline}
where $W(x,y)$ is the Gaussian window and the subscripts denote derivatives. The Gaussian window may be a representation of a composition of two separable kernels with binomial coefficients. Iterating through the system can yield even better results. It means that the retrieved offset is used to determine a new window in the second image from which the window in the first
image is subtracted, while $I_{t}$ is calculated.

\section{Pedestrian tracking using OpenCV software}
To determine pedestrian dynamics in the mosque of Mecca, with millions of people performing their rituals, we chose to use the OpenCV tools from Intel. This section describes the structure, operation, and functions of the open source computer vision library (OpenCV) for the Intel Corporation architecture \cite{Intel1999}. The OpenCV library is mainly used for real time computer vision. Some example areas are
human-computer interaction (HCI); object identification, segmentation, and
recognition; face recognition; gesture recognition; motion tracking, ego motion, and motion understanding; structure from motion (SFM); and mobile robotics.

\subsection{Results}  
Image sequences were obtained from videos collected by hd-cameras at the Hajj-2009. 

\begin{figure*}[ht]
\begin{center}
\includegraphics[width=1.0\linewidth]{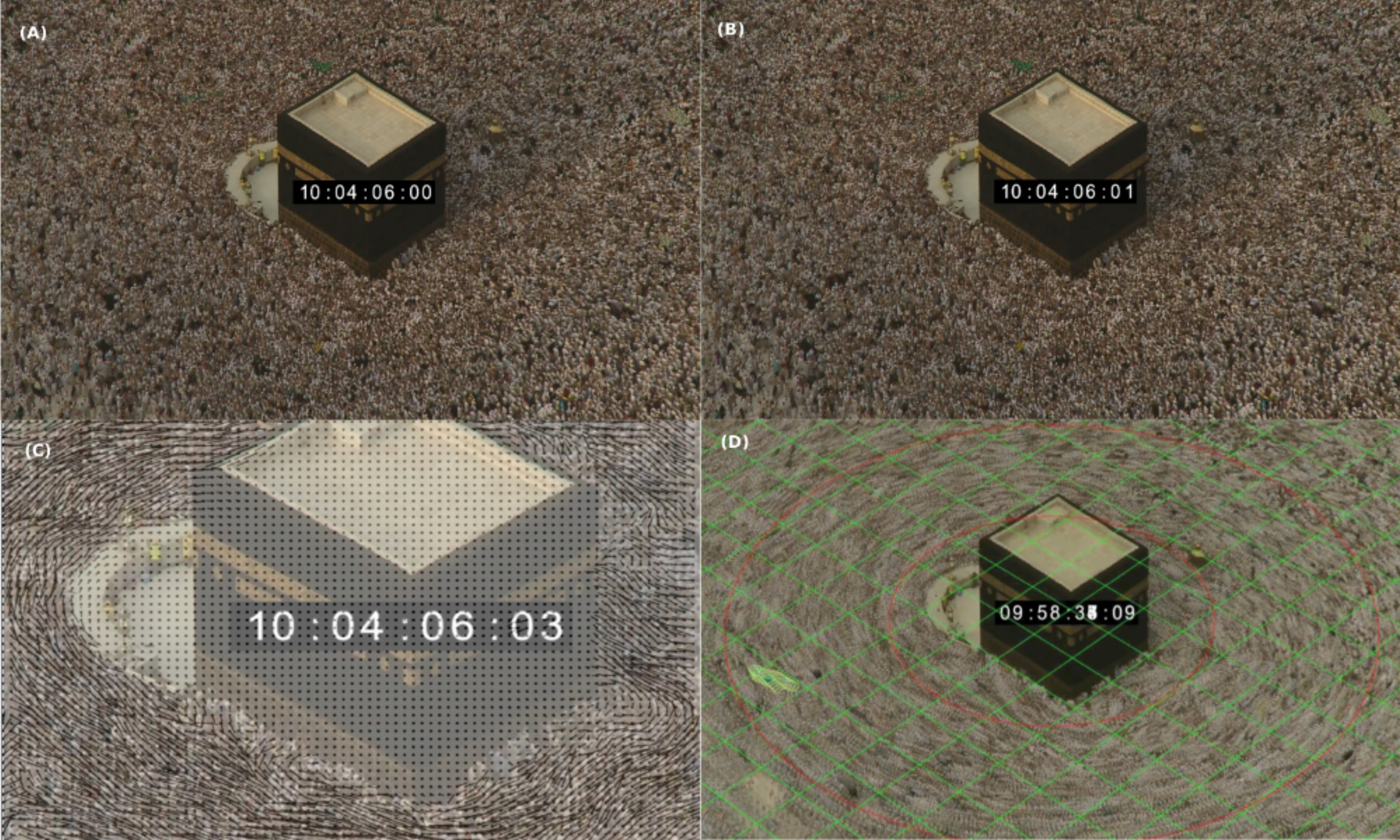}
\end{center}
\caption{Picture Analysis through Optical Flow Tools. Pedestrian flow walking around the Kaaba in the Haram in Mecca: (A) and (B)  illustrate two consecutive images in the sequence; each one of them consists of 1920$\times$1080 images pixels, (C) shows a set of velocity vectors obtained by the Lucas and Kanade technique computing at each point 
of a 64$\times$64 grid centred on the 1920$\times$1080 pixels, (D) shows the echo effect.}
\label{fig:echo1}
\end{figure*} 

The flow fields in figure \ref{fig:echo1} (C) show examples of the rotational movement of pilgrims around the Kaaba. We can clearly observe a kind of oscillation in the pilgrim paths around the Kaaba, this oscillation is caused by the shock-wave effect as a result of the repulsive forces between pedestrians in high density crowd dynamics. This was generated by applying the algorithm to every eighth pixel position on a pair of 1920$\times$1080 pixel images of a surface similar to figure \ref{fig:echo1} (A, B). 
The rotation field in \ref{fig:echo1}(C) was obtained by rotating pedestrian displacement in the Mataf area near the Kaaba wall. Through OpenCV tracking tools, it is possible to see that the movement around the Kaaba is not a perfect circular movement. The tracking of a simple individual in the pilgrim stream indicates some oscillation movement around the main path of the individual. These phenomena are due to the huge physical repulsive and attractive forces influencing the pedestrians movement. The pedestrian motion disturbance caused by high density crowd movement was also clearly visible in our pedestrian tracking on the Mataf area (see fig.  \ref{fig:pilgrimspaths-2}). This finding agrees with the video observation on the piazza of the Haram. For verification of the PedFlow approach we compare our simulation results with those of the optical flow. With help of this approach a lot of phenomena  (like the lane formation, oscillation effect and edge effect) can be seen, showing that our simulation reproduces a part of the reality. Therefore we stress that optical flow methods are very efficient for image analysis, structural analysis, image recognition, motion analysis and object tracking.  

The above mentioned techniques can be helpful for the validation and verification of simulation tools but is not sufficient, since there are many effects affecting the credibility of this method, for example: ambiguity, aliasing, and the aperture effect. One of these effects the 'aperture problem', has been extensively detailed in optical flow literature \cite{Horn1981}. However, the other two
short-comings (ambiguity, aliasing) are discussed to a lesser extend. Computer scientists and algorithms developer are working to resolve this problem so that the programs can take into account the
three points.  

\subsection{Echo Effect After Effects}
'Adobe After Effects \textregistered'\footnote{\href{http://www.adobe.com/de/products/aftereffects.html}{http://www.adobe.com/de/products/aftereffects.html}} is a digital motion graphic and composition software published by Adobe Systems \textregistered, used in the post-production process of film and television production. It is used for creating motion graphics and visual effects. 

After Effects helps us to understand the fluidity of the pedestrian flow and the density waves observed in the video recording during the rush hour on the Haram. These density waves are generated by huge pedestrian forces that propagate with the help of body contact through a crowd.

\begin{figure*}[ht]
\begin{center}
\includegraphics[width=1.0\linewidth]{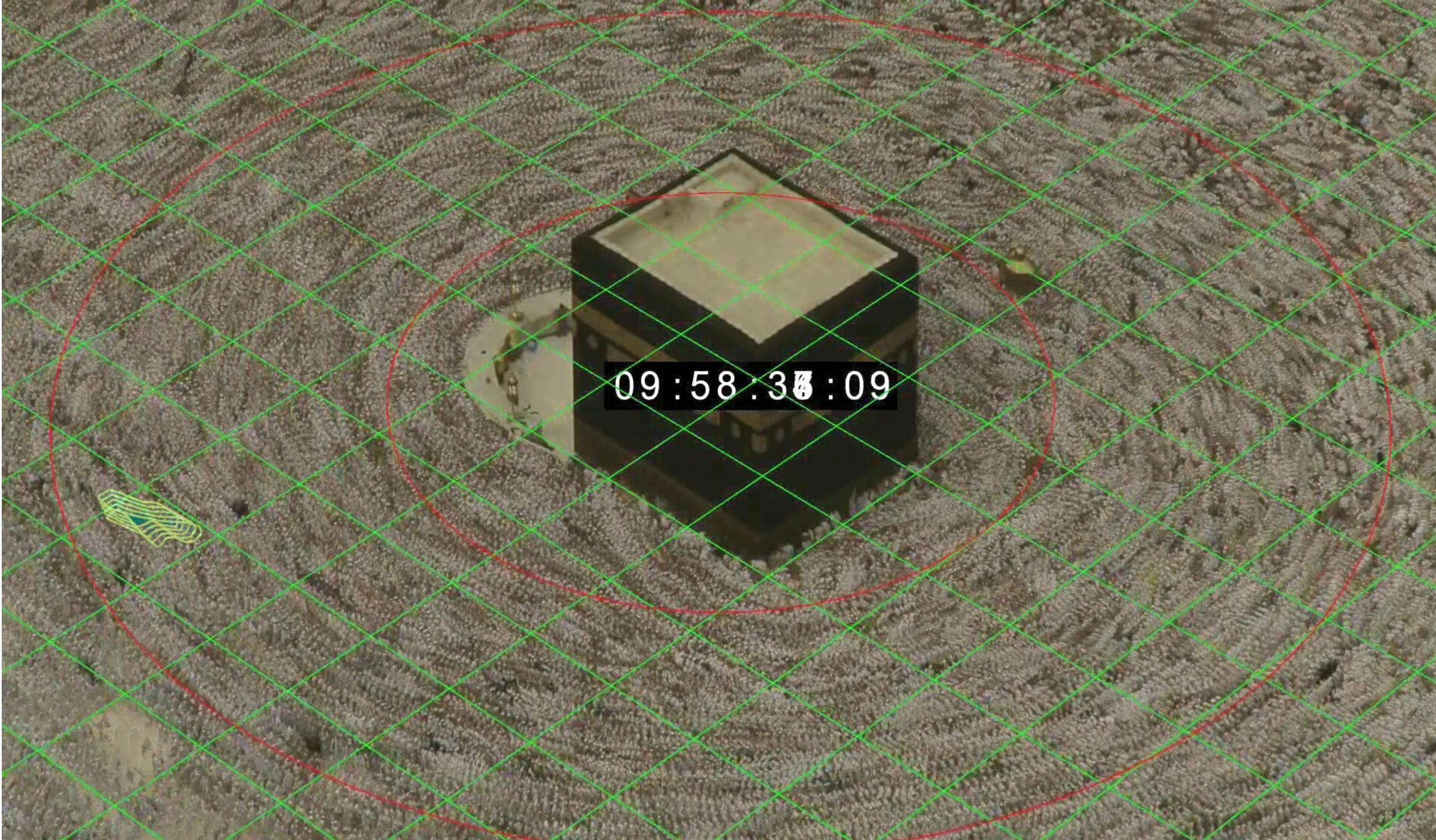}
\end{center}
\caption{Density waves (echo-effect): gray cloud-like structures near the Kaaba.}
\label{fig:echo-1}
\end{figure*}

\begin{figure*}[ht]
\begin{center}
\includegraphics[width=1.0\linewidth]{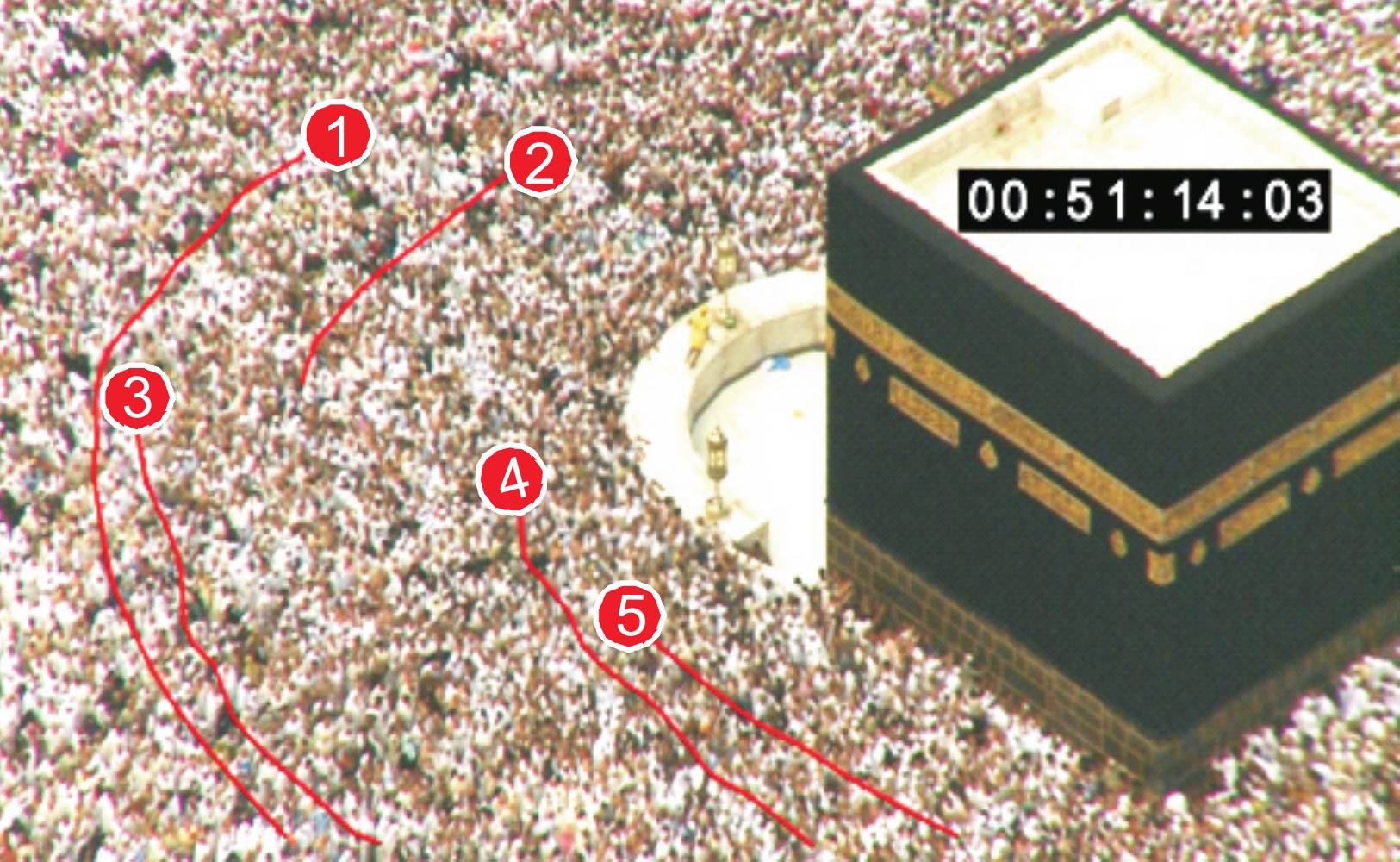}
\end{center}
\caption{Pilgrims paths. With a new computer
algorithm developed during this research, the trajectories or movements of pedestrians across the 
infrastructure over time are determined. Microscopic pedestrian fields require large 
amounts of trajectory data of individual 
pedestrians. Every red solid curve corresponds to one pedestrian trajectory. The oscillation in the pilgrims paths results from the huge pedestrian forces acting on every individual in the crowd.}
\label{fig:pilgrimspaths-2}
\end{figure*}

In figure \ref{fig:pilgrimspaths-2} we show the path of individuals within the crowd. One clearly recognizes that the movement around the Kaaba is not a circle movement. The tracking of a single individual in the pilgrims stream indicates some oscillation movement around the main path of the individual, as already mentioned it is caused by the physical repulsive and attractive forces acting on the individual. Physical forces become important when an individual comes into physical contact with another individual/obstacle.  
When a local density of 6 persons per square meter is exceeded, free movement is impeded and local flow decreases, causing the outflow to drop significantly below the inflow. This causes a higher and higher compression in the crowd, until the local densities become critical in specific places on the Mataf platform. This technique has helped to demonstrate the density waves, and that the movement around the Kaaba is not circular but disturbed movements caused by this density shock waves. The disturbance in the path of the pilgrims is generated by the enormous contact forces that come into play in this region especially near the Kaaba (see fig.\ref{fig:echo-1}). These waves appear in figure \ref{fig:echo-1} as gray structures around the Kaabe.
 
\subsection{Discussion}  
\begin{figure*}[ht]
\begin{center}
\includegraphics[width=1.0\linewidth]{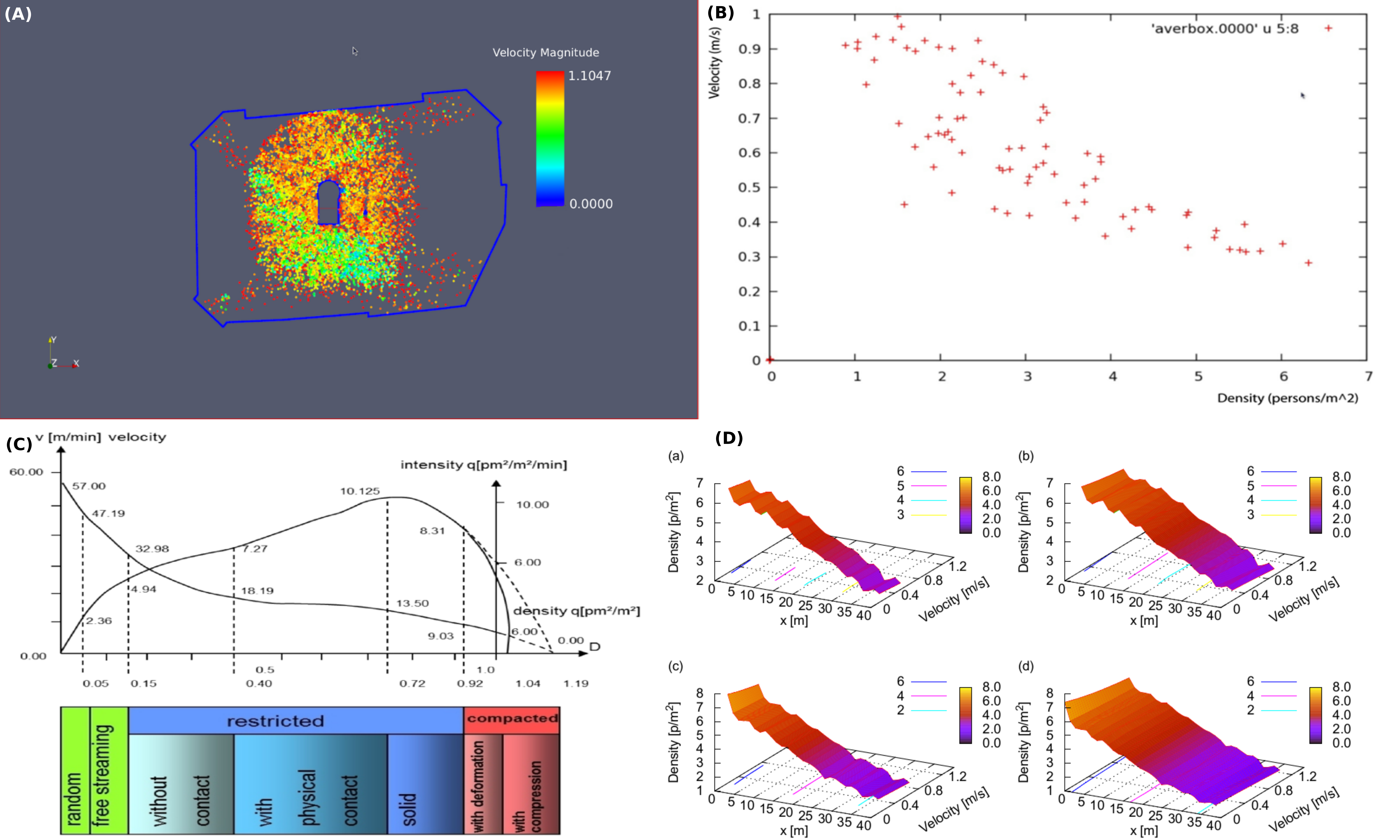}
\end{center}
\caption{(A) A snapshot of tawaf simulation results made by PedFlow with velocity index, the blue color indicates the pedestrian stand still while the red color indicates their maximal walking speed; (B) velocity-density diagram: PedFlow simulation results; (C) Predtetschenski and Milinski fundamental diagram; (D) velocity-density distribution as a function of the distance from the Kabaa wall, the curves (b, d) illustrate the simulation results while the curves (a, c) show the statistical results and indicate the density behaviour on the Mataf area at different times during the day, obtained by a calculation according to the Predtetschenski and Milinski fundamental diagram \cite{dridi2014}.}
\label{fig:verification3}
\end{figure*}

We have used a multi-stage validation method. 
One of the most important parameters was verified, the pedestrian density distribution on the Mataf area as a function of the position $\vec{r}$ and velocity $\vec{v}$.
It served in a first step of a comparison of the simulation density results with the observed density behaviour on the Mataf area at different times during the day, before and after the prayer. The maximum registered density obtained by the statistical method was 7 to 8 persons/m$^{2}$. One can clearly recognise the similarity between the statistical data and the results given by the simulation, which can reach 7 persons/m$^{2}$, especially in the congested area (see fig. \ref{fig:verification3} (A) and (B)). 

From observation of the Mataf it is well-known that the area indicating the beginning and the end of the Tawaf is the area most congested and accumulated by pilgrims, (see fig.\ref{fig:verification3}) (D). This phenomenon is obviously reproduced by the simulation (see fig. \ref{fig:verification3} (A)). This area appears in the picture on the right lower corner of the Kaaaba, known as black stone corner where the observed pilgrim density reached over 9 persons/m$^{2}$. All statistical results illustrating the density distribution at the Mataf area at different time intervals are demonstrated in the paper \cite{dridi2014}.

The second step of the multi-stage verification was to compare the velocity-density diagram made by the simulation with all well-known fundamental diagrams. According to Predtetschenski and Milinski the average walking speed depends on the the walking facility and the local density which can reach 9 persons/m$^{2}$ \cite{Predtetschenski-Milinski1969}. In figure \ref{fig:verification3} one clearly recognizes density waves with maximum density near the Kaaba wall. There the average local density can reach a critical value of 7 to 8 persons/m$^{2}$. In the congested area the local density increases with significant dropping in the pedestrian velocity. The average local speed $\vec{v} (\vec{r}, t)$  as a function of the local density $\rho(\vec{r}, t)$ made by the simulation is compared with the Predtetschenski and Milinski densities in figure \ref{fig:verification3} (C). Our own data is shown as red crosses in figure \ref{fig:verification3} (B). Moreover the analysis of the data of the Mataf area showed that a reduction of the available navigation space is responsible for the speed reduction and the density increase. The small deviation in pedestrian walking speeds at lower density can be explained by the fitness level of the pedestrian. Through this comparison two phenomena are clearly demonstrated, the density effect in the Mataf area and the edge effects: the edges of a crowd move faster than the center of the crowd. This phenomenon was clearly demonstrated in the statistical results shown in figure \ref{fig:verification3} (D).

Comparing the results of PedFlow with results of other models in the simulation of a special cases like the Jamarat bridge (see fig. \ref{fig:Djamarat2}), showed that the critical points in the Jamarat facility made by microscopic simulations with Myriad \cite{Still2003}  are the same critical points exhibited by the PedFlow microscopic simulations of the Jamarat bridge.

\section{Conclusions and recommendations}
For people working in software development and simulation program evolution the verification and validation of the model is a vital procedure to make sure that the tools apply to reality. The validation of the simulation program ensures the users and decision makers that the simulation results are credible and applicable in the development of the project. Moreover Turing and face validity tests contribute to progressive optimization of the program.
The Turing test is a successful method comparing the real world with the simulation output. The output data obtained by the simulation can be presented to people attending the same project and working with the same tools knowledgeable about the system in the same exact format as the system data.  
The discussion between the experts about the deviation of the simulation and the system outputs can be helpful to validate the program, their explanation of how they did that should improve the model.

The opinion of the project member and model user for development, progress and verification of the simulation tools is very important. This method will be referred to as face validation. Face validation is necessary to identify the behaviour of the simulation system under the same simulation condition. A preliminary examination of the model one can deduce that this method is useful, necessary, but not sufficient.

In this paper we have discussed verification and validation of microscopic simulation models. Different approaches and methods for deciding verification and validation of the model development process have been presented, as have been various validation techniques. 
 
As a practical example, the Haram Mosque in Mecca and the Jamarat Bridge in Saudi Arabia were used for high density crowd simulation: the huge number of pilgrims cramming the bridge during the pilgrimage to
Mecca gave rise to serious pedestrian disasters in the nineties. Moreover, the analytical and numerical study of the qualitative behaviour of human individuals
in a crowd with high densities can improve traditional socio-biological investigation
methods. 

For obtaining empirical data different methods were used, automatic and manual methods. We have analysed video recordings of the crowd movement in the Tawaf in Mosque/Mecca during the Hajj on the 27th of November, 2009. We have evaluated unique video recordings of a
105$\times$ 154 m large Mataf area taken from the roof of the Mosque, where upto 3 million Muslims perform the Tawaf and Sa'y rituals within 24 hours.

For the validation and calibration of the simulation tools, different methods were used. 
\begin{itemize}
\item Comparison of the simulation result with the video recording.
\item Comparison with other models: Different results (e.g., outputs) of the simulation model being validated, and compared with the results of other models. For example, (1) simple cases of a simulation model were compared with well-known results of analytic models, and (2) the simulation model were compared with other simulation models that have been validated.
\item Parameter Variability - Sensitivity Analysis: Applying this  technique one can determine the behaviour of the model or simulation output, using different input values.
\item A comparison with Optical Flow results was also carried out.
\end{itemize}
At medium to high pedestrian densities, the techniques used in PedFlow can produce realistic crowd motion, with pedestrians moving at different speeds and under different circumstances, following believable trails and taking sensible avoidance action. 

\section{Acknowledgements}
I would like to express my sincerest thanks and gratitude to Prof. Dr. G. Wunner for a critical reading of the 
manuscript, for his important comments and suggestions to improve the manuscript. Many thanks to Dr. H. Cartarius for his support during writing this work.

\bibliographystyle{unsrt}

\end{multicols}
\end{document}